\RequirePackage{fix-cm}
\documentclass[smallextended]{svjour3}       % onecolumn (second format)
\smartqed  % flush right qed marks, e.g. at end of proof
\usepackage{graphicx}
\usepackage{amsmath,amssymb}
\usepackage{mathptmx}      % use Times fonts if available on your TeX system
\usepackage{color}
\begin{document}

\title{Metamaterial lens of specifiable frequency-dependent focus and adjustable aperture for electron cyclotron emission in the DIII-D tokamak%\thanks{Grants or other notes
}

\titlerunning{Metamaterial lens of specifiable frequency-dependent focus}        % if too long for running head

\author{K. C. Hammond         \and
              W. J. Capecchi  \and
              S. D. Massidda  \and
              F. A. Volpe  %etc.
}


\institute{K. C. Hammond \at
              Columbia University, New York, NY, USA
              %\email{f@example.com}           %  \\
           \and
           W. J. Capecchi \at
           University of Wisconsin, Madison, WI, USA 
           \and
           S. D. Massidda \at
           Columbia University, New York, NY, USA
           \and
           F. A. Volpe \at
           Columbia University, New York, NY, USA \\
           \email{fv2168@columbia.edu}
}

\date{Received: date / Accepted: date}

\maketitle

\begin{abstract}
Electron Cyclotron Emission (ECE) of different frequencies originates at different locations in non-uniformly magnetized plasmas. For simultaneous observation of multiple ECE frequencies from the outside edge of a toroidal plasma confinement device (e.g. a tokamak), the focal length of the collecting optics should increase with the frequency to maximize the resolution on a line of sight along the magnetic field gradient. Here we present the design and numerical study of a zoned metamaterial lens with such characteristics, for possible deployment with the 83--130 GHz ECE radiometer in the DIII-D tokamak. The lens consists of a concentric array of miniaturized element phase-shifters. These were reverse-engineered starting from the desired Gaussian beam waist locations and further optimized to account for diffraction and finite-aperture effects that tend to displace the waist. At the same time we imposed high and uniform transmittance, averaged over all phase-shifters. The focal length is shown to increase from 1.37 m to 1.97 m over the frequency range of interest, as desired for low-field DIII-D discharges (B = -1.57 T). Retracting the lens to receded positions rigidly moves the waists accordingly, resulting in a good match---within a fraction of the Rayleigh length---of the EC-emitting layer positions at higher fields (up to B = -2.00 T). Further, it is shown how varying the lens aperture might move the waists ``non-rigidly" to better match the non-rigid movement of the EC-emitting layers with the magnetic field. The numerical method presented is very general and can be used to engineer any dependence of the focal length on the frequency, including zero or minimal chromatic aberration.
\end{abstract}

%%% __________________________ Introduction _____________________________ %%%
\section{Introduction}
\label{sect:introduction}

The ability to direct millimeter waves into and out of a magnetically confined plasma is of great value in plasma research. One principal application is the detection of electron cyclotron emission (ECE) to infer the electron temperature profile.~\cite{bornatici1983} This inference is permitted by two properties of tokamak plasmas: (1) The magnetic field in the plasma is non-uniform, scaling like $1/R_\text{maj}$, where $R_\text{maj}$ is the distance from the axis of symmetry of the torus. (2) The electron temperature is proportional to the radiative temperature, a consequence of the tokamak plasma being a blackbody emitter in the ECE range of frequencies. 

The ECE frequency distribution over $R_\text{maj}$ is determined primarily by the strength of the toroidal magnetic field at a given $R_\text{maj}$, with corrections for Doppler and relativistic broadening \cite{clark1983}. The magnetic field (and, therefore, the electron cyclotron frequency) in a tokamak is roughly inversely proportional to the major radius; hence, electrons at smaller (larger) $R_\text{maj}$ emit ECE of higher (lower) frequency. The radiation can be spectrally analyzed by an ECE diagnostic, typically located on the outside edge tokamak where the magnetic field is weakest. To achieve the best spatial resolution, therefore, the diagnostic should receive the radiation through a focusing element whose focal length increases with frequency---in other words, the element should exhibit reverse chromatic aberration (RCA), a frequency dependence opposite to what is normally encountered in natural materials.

Incorporating an optic with RCA would have the potential to improve the quality of ECE detection on the DIII-D tokamak \cite{luxon2002}. In the current setup, ECE from the plasma is reflected off an ellipsoidal mirror on the low-field side of the tokamak, after which it is received by a scalar horn antenna connected to a 40-channel radiometer. The mirror has essentially the same focal length at all frequencies \cite{ellis2006}. The frequencies detected by the radiometer, however, are emitted from a range of major radii that differ by up to 0.85 m and vary with toroidal field strength. Replacing the mirror with an RCA lens would enable higher spatial resolution. Additionally, moving the RCA lens would move its foci in response to the changes in toroidal field strength.

The utility of a lens with a specifiable dependence of focal point on frequency may extend beyond tokamak ECE detection. For example, the lens could be useful for the detection or irradiation of a moving target for which the particular frequency is unimportant. In such cases, the location of the focus may be adjusted rapidly simply by tuning the frequency of the receiver or source.

Techniques to control chromatic aberrations are not new. The oldest example is the achromatic doublet, a combination of a convergent and divergent lens of different materials and therefore different amounts of dispersion. While the focal length of each lens is a monotonic function of frequency $f$, the focal length of a doublet is approximately quadratic in $f$; it matches a desired focal length $\ell$ exactly at two frequencies and approximately in a range around them. The concept is readily generalized to ``apochromatic" triplets or ``superachromatic" quadruplets of lenses, where $\ell$ takes a certain value at three or four frequencies. That value is typically the same, for the sake of minimizing chromatic aberration \cite{born2006}. This technique can, in theory, be applied to produce the reverse chromatic aberration desired for our ECE optics. It would have significant disadvantages, however. In particular, (1) its degrees of freedom are limited to two per lens (the focal length $\ell$ for a certain $f$ and the material, which indirectly fixes the dependence at other frequencies, $\ell(f)$) and (2) there is a practical limit to the maximum number of lenses that can be arrayed or stacked together. 

A metamaterial gradient-index lens approach, on the other hand, would avoid both of these limitations. Metamaterial lenses, typically thinner than one wavelength, consist of hundreds of microscopic unit cells whose dimensions can be independently specified to achieve certain optical properties, such as negative refractive indices \cite{schurig2005,smith2005,smith2006,cui2009}. One approach to developing such metamaterials is to design a miniaturized-element frequency selective surface (MEFSS) \cite{behdad2007,hua2012}. The unit cells of such materials typically have sub-wavelength dimensions, allowing for the fabrication of surfaces whose filtering properties vary smoothly on length scales appropriate for lens applications. One application, which we have investigated previously, is to engineer an MEFSS-based lens exhibiting RCA with a specific $\ell(f)$ dependence \cite{capecchi2012}. This paper describes in detail our methods to tailor the focusing characteristics of an MEFSS-based lens to meet the requirements for ECE detection at DIII-D.

\begin{figure}
\includegraphics[width=7cm]{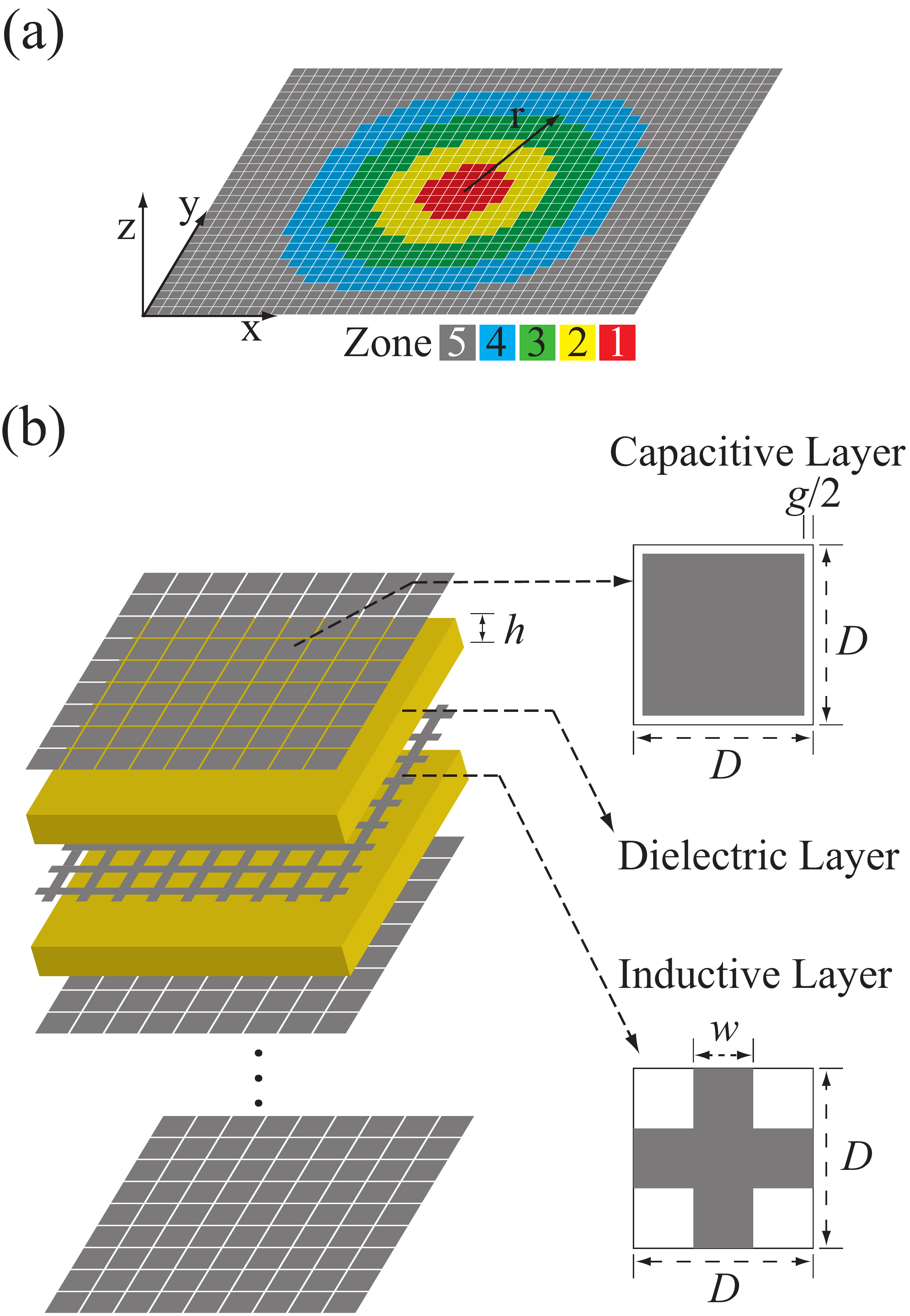}
\caption{(a) Schematic illustrating the partitioning of the unit cells (squares) of the metamaterial lens into concentric zones (differentiated by color). Each zone is defined by the phase-advance that its component unit cells impart to incident radiation. The schematic portrays a lens with 5 zones, whereas this paper presents a design with 83 zones. (b) Schematic illustrating the layered structure of an MEFSS, which consists of alternating layers of capacitive patches and inductive mesh separated by dielectric material. The parameters $g$ and $w$ associated with each unit cell are defined in the insets. (c) Model of an MEFSS as an AC transmission line, illustrating how the lens behaves as a spatial phase shifter. For a more detailed discussion, see Ref.~\cite{al-joumayly2010}. Adapted from Ref.~\cite{capecchi2012} with permission.}
\label{fig:uc_schematic}
\end{figure}

Although single lenses made of natural materials are generally constrained to impart greater dispersion to waves with higher $f$ (\textit{i.e.}, exhibit traditional chromatic aberration), metamaterials, in general, do not face this constraint \cite{silveirinha2009}. In particular, RCA has recently been observed experimentally in certain MEFSS-based lenses designed for microwave frequencies \cite{al-joumayly2011}. The surfaces in these lenses consist of alternating layers of square metal patches (capacitive layers) and wire grids (inductive layers) (Fig.~\ref{fig:uc_schematic}b). Altering the dimensions of the unit cells will affect the phase-advance of electromagnetic radiation transmitted through them (see Ref.~\cite{al-joumayly2010} for a detailed discussion on wave propagation through these materials). An MEFSS whose unit cell parameters (and, by consequence, spatial phase-advance) vary properly as a function of distance from the transverse axis will exhibit lens-like behavior. 

Previous lenses of this kind have been designed by partitioning an MEFSS into a set of discrete annular zones concentric with a perpendicular axis (Fig.~\ref{fig:uc_schematic}a), with each zone associated with a certain phase-advance \cite{al-joumayly2011,capecchi2012}. Outer zones (i.e. those with larger annular radii) impart greater phase-advances than those near the axis, such that an incident collimated beam of light that passes through a zoned MEFSS will undergo a transformation in its radial phase profile and taper to a focal point a certain distance beyond the lens. Owing to the high number of degrees of freedom in the design of each unit cell, it is possible to identify a geometrical configuration having specified focal lengths for certain frequencies---including a configuration with RCA, as we have previously shown numerically in the 8-12 GHz band \cite{capecchi2012}. The appropriate unit cell parameters for each zone were chosen based on the results of simulations of radiation passing through single unit cells.  

In the present paper, we have expanded upon the methods of Ref.~\cite{capecchi2012} to design an RCA lens optimized for ECE in the DIII-D tokamak. The lens has an aperture diameter of 0.30 m, equal to the diameter of the viewing port onto which it might be installed. 

The increase in aperture from Ref.~\cite{capecchi2012} will necessitate a greater range of phase-advances from the unit cells. One way of accommodating this increased range is to increase the lens order, defined as the number of capacitive layers (equivalently, the number of inductive layers plus one). From the point of view of the unit cell, adding extra layers amounts to stacking on extra spatial phase shifters. In this way, discrepancies in phase-advance between single layers can be magnified, allowing for greater variations in phase-advance overall between lens zones. Practical considerations put limits on lens order, however: each added layer introduces new absorptive losses and increases fabrication costs. For this paper we have chosen to work with a 10th-order lens to strike a balance between these opposing considerations.

For improved accuracy in our determination of the desired spatial phase-advances for the lens unit cells, we have used Gaussian optics rather than geometric optics, which is more appropriate for the frequency and length scales of our experiments. These determinations were further optimized through corrections for the effects of a finite lens aperture diameter. We have also improved our method for selecting optimal unit cell dimensions based on simulation data. 

Sec.~\ref{subsect:phase-advances} describes how we determined the desired phase advance for each unit cell on the metamaterial lens. Sec.~\ref{subsect:database} outlines our use of simulation software to build a database of candidate unit cells. Sec.~\ref{subsect:optimization_motivation} provides motivation for the algorithm we developed for choosing the unit cells from the database what would for a lens that best matched the ideal behavior; Sec.~\ref{subsect:lens_parameters} presents the algorithm. Sec.~\ref{subsect:numerical_optimization} describes further measures taken to fine-tune the unit cell dimensions chosen from the database. Sec.~\ref{subsect:predicting_performance} outlines our computations for quantitatively predicting the performance of the optimized lens.  Sec.~\ref{subsect:non-ideal-properties} presents the results of electric field computations that control for lens aperture, which informed our optimization procedures. Sec.~\ref{subsect:parameter_optimization} presents the unit cell parameters chosen through our optimization procedure. Sec.~\ref{subsect:predicted_performance} presents the results of computations that predict the performance of the optimized lens.

%%% ______________________________ Methods ______________________________ %%%
\section{Numerical Methods}
\label{sect:methods}

%%% _______________ Methods: Determining the desired phase advances _____________ %%%
\subsection{Determining the desired phase advances}
\label{subsect:phase-advances}

\begin{figure}
\includegraphics[width=7cm]{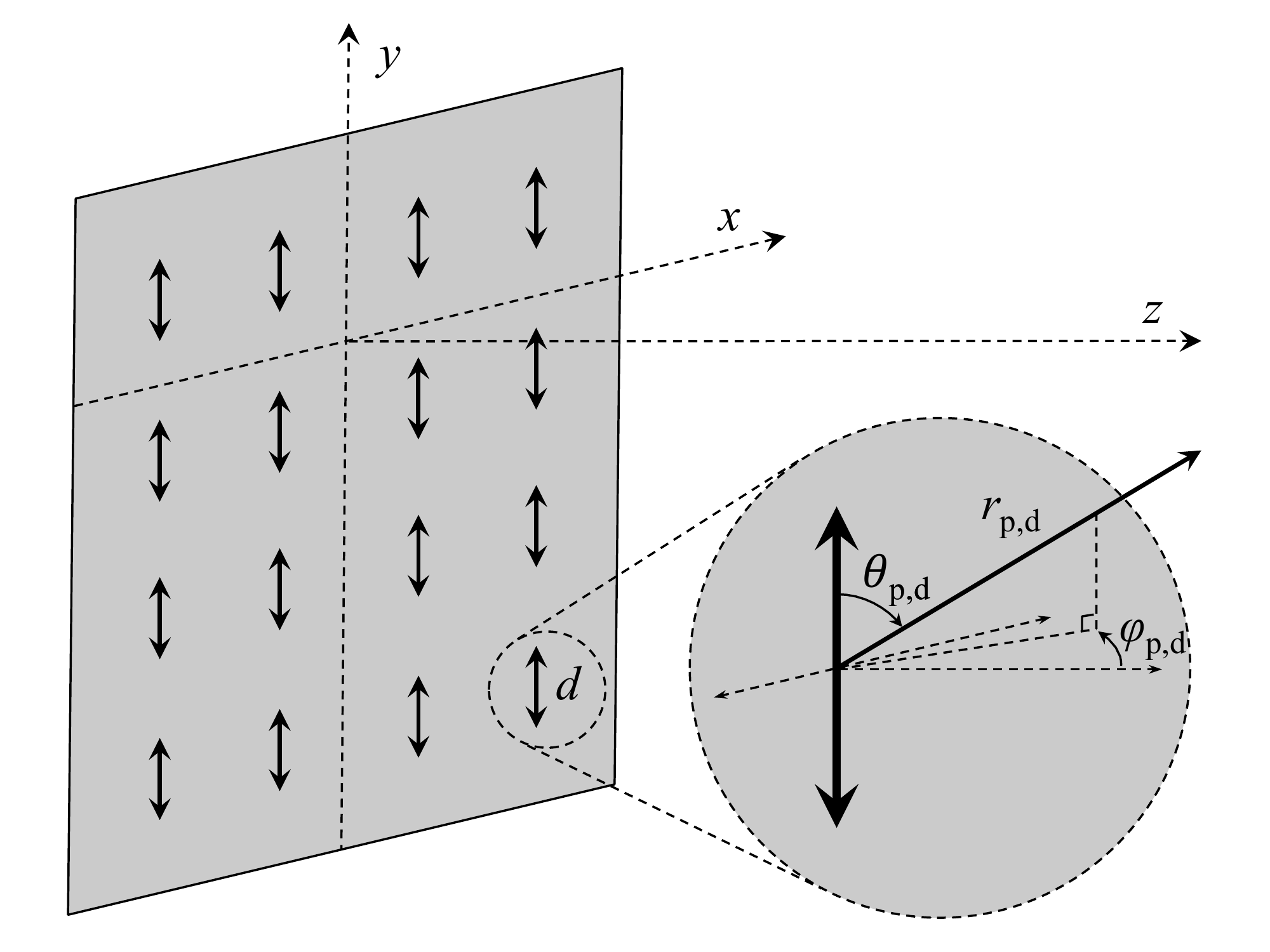}
\caption{Schematic setup for field intensity computations. The lens is in the $xy$-plane, with each dipole polarized parallel to the $y$-axis (the choice of polarization is arbitrary). The electric field at a point $p$ (in the $xyz$ coordinate system) is determined by computing the coherent sum of the contributions from each dipole $d$. Each dipole represents a single unit cell of the metamaterial lens.}
\label{fig:computation-schematic}
\end{figure}

In the regime of Gaussian optics \cite{goldsmith1998}, waves are treated as a superposition of Gaussian modes (\textit{i.e.}, solutions to the paraxial wave equation). For waves that propagate along an axis perpendicular to the lens (the $z$-axis in Fig.~\ref{fig:computation-schematic}), the electric field of each Gaussian mode will be of the form

\begin{equation}
\label{eqn:gaussian_mode}
E \propto \exp\left[-\frac{r^2}{s(z)^2} - ikz - i\Phi(z,r,f) + i\Phi_0\right],
\end{equation}

\noindent where $r$ is the distance from the propagation axis, $z$ is the distance along the axis, $s(z)$ is a characteristic transverse radius for the beam, $k$ is the wavenumber, $\Phi_0$ is an arbitrary phase offset, and $\Phi(z,r,f)$, hereafter the \textit{phase profile}, is

\begin{equation}
\label{eqn:phase_profile}
\Phi(z,r,f) = \frac{\pi f}{c R(z,f)}r^2,
\end{equation}

\noindent where $f$ is the frequency and $R(z,f)$ is the radius of curvature. 

We will here define the focal length $\ell$ at frequency $f$ of a lens as the distance of the beam waist of an outgoing wave of frequency $f$ from the lens assuming the incoming wave had a uniform phase profile (infinite $R$) at its point of incidence with the lens. (This requires the assumption that the outgoing wave is a Gaussian mode with a well-defined beam waist; in practice, however, the output will be a superposition of modes.) Note that this is slightly different from geometric optics, which defines the focal length as the convergence point for parallel incident rays after refraction \cite{goldsmith1998}. To achieve this output, the lens must convert the phase profile of the incoming wave (assumed to be uniformly zero) to the profile $\Phi(\ell,r,f)$.

In our work, we begin with a set of focal lengths $\ell$ (Table \ref{table:benchmark_frequencies}) that the lens must have for a corresponding set of frequencies $f$. This determines the phase profile $\Phi(\ell,r,f)$ that the lens must impart to the outgoing wave; equivalently, it determines the radius of curvature $R(\ell,f)$ associated with the phase profile. Thus, in theory, the lens needs to impart a phase-advance $\phi$ to the incoming wave (assumed to be a plane wave with $R = \infty$ at incidence) at radial distance $r$ by a phase equal to $\Phi(\ell,r,f)$. One may accomplish this by partitioning the lens's unit cells into concentric annular \textit{zones} such that a unit cell in the $n^\text{th}$ zone (of annular radius $r_n$) imparts a phase-advance equal to the desired phase profile plus an arbitrary constant: $\phi(n,f) = \Phi(\ell,r_n,f) + \Phi_0$.

\begin{table}
\begin{tabular}{|c|c|}
     \hline
     Frequency $f$ (GHz) & Focal Length $\ell$ (m) \\ \hline \hline
     83.5 & 1.249 \\ \hline
     92.5 & 1.466 \\ \hline
     101.5 & 1.671 \\ \hline
     110.5 & 1.830 \\ \hline
     119.5 & 1.959 \\ \hline
     129.5 & 2.082 \\ \hline
\end{tabular}
\caption{List of benchmark frequencies $f$ for ECE emission and the corresponding desired focal lengths $\ell$. The desired focal lengths are based on the emission locations in the DIII-D tokamak.}
\label{table:benchmark_frequencies}
\end{table}

One important consideration that we have thus far omitted in the discussion is the finite aperture of the lenses. Setting $\phi = \Phi$ would yield the desired $\ell$ if the aperture were much wider than the beam, but this cannot be assumed with our aperture radius of 15 cm. In practice, to achieve a desired focal length, the phase-advance $\phi$ of each zone must correspond to a slightly greater ``adjusted" radius of curvature $R_\text{adj}$ than that of the desired output Gaussian mode.

To determine these corrected radii of curvature, we modeled the metamaterial lens as an array of radiating electric dipoles (Fig.~\ref{fig:computation-schematic}), each of which corresponded to a single unit cell of the MEFSS. Such a dipole array is commonly used to represent a lens consisting of discrete phased elements, with experimental results showing good agreement with the computations \cite{al-joumayly2011,tetienne2011}. Field computations using this setup are much faster than numerical solutions of electromagnetic waves propagating through a simulated metamaterial lens.

As with the unit cells in the metamaterial lens, the dipoles in the computations were assigned to annular zones based on their distance from the beam axis. The dipoles in each zone were given an amplitude and phase corresponding to a Gaussian mode (Eq.~\ref{eqn:gaussian_mode}) evaluated at frequency $f$ with $s = 9.8$ cm (chosen such that 99\% of the beam energy would pass through the aperture of radius 15 cm) and $\Phi$ initially equal to the phase profile $\Phi(\ell,r_n,f)$ corresponding to the desired focal length $\ell$.  The ``actual" focal length $\ell_{\text{act}}$ of this setup was determined by finding the point of maximum field intensity along the $z$-axis. To determine the appropriate phase-advance $\phi(n,f)$ for the $n$th zone at frequency $f$, the radius of curvature $R$ associated with $\Phi$ was tweaked until $\ell_\text{act}$ converged to $\ell$. This (frequency-dependent) adjusted radius of curvature is $R_\text{adj}$ as defined above; thus the phase advance of zone $n$ at frequency $f$ should be

\begin{equation}
\label{eqn:phase_shift}
\phi(n,f) = \frac{\pi f}{c R_\text{adj}(f)}r_n^2 + \phi_0(f)
\end{equation}

\noindent where $\phi_0(f)$ plays the role of $\Phi_0$ in Eq.~\ref{eqn:gaussian_mode}.

Note that there is an inherent degree of freedom in the phase-advances of the unit cells: as long as the phase-advances of each zone $\phi(n,f)$ vary correctly \textit{relative to each other}, their \textit{absolute} phase-advance is arbitrary; \textit{i.e.}, $\phi(n,f)$ may vary by an arbitrary constant $\phi_0(f)$ (which may vary with frequency in general but must be the same for all zones $n$ at a given $f$). This has important ramifications for the optimization process, discussed in \ref{subsect:optimization_motivation}. For now, we define the \textit{relative phase advance} $\Delta\phi$, an auxiliary quantity corresponding to the difference in phase advance of zone $n$ with that of zone 1 (the innermost zone):

\begin{align}
\label{eqn:Delta_phi}
\begin{split}
\Delta\phi(n,f) ~&\equiv~ \phi(n,f) - \phi(1,f) \\
&=~ \frac{\pi f}{cR_\text{adj}(f)}(r_n^2 - r_1^2)
\end{split}
\end{align}

%%% _______ Methods: Geometrical scans: building a database of unit cell responses _____ %%%
\subsection{Geometrical scans: building a database of unit cell responses}
\label{subsect:database}

With the desired relative phase-advances $\Delta\phi(n,f)$ at hand, the next task was to determine a set of unit cell parameters for each zone whose spatial phase-advances would match $\Delta\phi(n,f)$. This was accomplished by building a large database of unit cells with varying internal dimensions. Each of these unit cells imparted a different phase advance to an oncoming wave and exhibited a different transmittance. Furthermore, both phase advance and transmittance were liable to vary with frequency within the same cell and such dependencies varied with the geometric properties.

Each unit cell (Fig.~\ref{fig:uc_schematic}b) was of 10th order with capacitive layers on both ends. The cells had square cross sections with 600 $\mu$m sides. Layers were separated by 509 $\mu$m of dielectric material. The two parameters scanned for the database were the capacitor gap $g$ (defined as twice the spacing between the edge of a capacitive patch and the unit cell border) and the inductor width $w$ (defined as the side length of the square hole in a unit cell of the wire grid) \cite{capecchi2012}. Each capacitive and inductive layer within a given unit cell had the same value of $g$ and $w$ (respectively). Values of $g$ ranged from 80 $\mu$m to 272 $\mu$m at intervals of 2 $\mu$m; $w$ ranged from 0 $\mu$m to 40 $\mu$m at intervals of 2 $\mu$m.

The phase and transmittance properties of a unit cell of given $g$ and $w$ were computed in frequency-domain CST Microwave Studio simulations. In each simulation, a wave packet was launched through a single unit cell with periodic boundary conditions. Transmittance $T$ and the difference in phase between launching and receiving ports (which we will define as $\delta\phi$) were computed for six benchmark frequencies (Table \ref{table:benchmark_frequencies}) corresponding to channels of the DIII-D ECE radiometer. The metal in the capacitive patches and wire grids was assumed to have the material properties of copper and had zero thickness. The dielectric material was assumed to be isotropic and linear.

Although the phase data $\delta\phi$ recorded by the solver for each unit cell does not contain information about precisely how many phase cycles a wave undergoes when passing from the simulated transmitter and receiver, it is still sufficient for the purposes of designing a lens. Since $\delta\phi$ of a unit cell is equal to its actual phase-advance $\phi$ plus an integer multiple of $2\pi$ radians, no information is lost about the unit cell's contribution to the interference effects of the lens.

%%% ______________ Methods: Motivation for optimization procedures ________________ %%%
\subsection{Motivation for optimization procedure exploiting the arbitrariness of zone 1}
\label{subsect:optimization_motivation}

The ultimate objective of the optimization is to select a set of unit cells which, when arranged in the zoned array, behaves as a lens with a specified set of focal lengths $\ell_i$ corresponding to the six benchmark frequencies $f_i$ specified above. The question now is how to choose the unit cells that most closely exhibit this behavior. This problem was explored in Ref.~\cite{capecchi2012}; here we present an improved method that relaxes some previously imposed constraints in order to achieve better agreement with the desired lens behavior.

The problem may be characterized as follows. One rudimentary approach would be to choose a random unit cell from the database with a certain $(g,w)$ and declare it to take the role of zone 1. (We will thus refer to its parameters as $(g_1,w_1)$.) This zone will impart a certain phase-advance $\phi(1,f_i)$ to each of the benchmark frequencies $f_i$ (recall that the first argument of $\phi$ is the zone number $n$). This phase-advance will then specify the desired phase-advances $\phi(n,f_i)$ for every remaining zone  $1<n\leq83$ as per Eq.~\ref{eqn:Delta_phi}. Then, for each $n$, one scans the database for the unit cells $(g_n,w_n)$ whose phase-advances are closest to $\phi(n,f_i)$.

This method may be improved upon by noting that a different choice of $(g,w)$ from the database as the zone 1 unit cell may yield a set of unit cells that better conforms to the desired lens behavior; thus, one can repeat the process in the above paragraph, using every zone from the database in turn as zone 1. This effectively tests $N$ hypothetical lenses if there are $N$ unique cells in the database, the best of which can then be chosen as the optimized lens prototype.

There is an additional improvement to this method. The above process requires implicitly that the zone 1 unit cell impart phase-advances $\phi(1,f_i)$ that conform exactly to one of the unit cells in the database (whereas the unit cells in the remaining zones have $\phi(n,f_i)$ that are only approximately equal to the exact $\phi(n,f_i)$ demanded by $\phi(1,f_i)$ via Eq.~\ref{eqn:Delta_phi}). One can obtain even more flexibility in the optimization by relaxing this requirement: instead of choosing unit cells from the database and using their calculated phase-advances as the exact set of phase-advances $\phi(1,f_i)$ for zone 1, one can instead use set of strategically chosen \textit{target functions} $\phi_t(1,f_i)$. This increased flexibility will theoretically lead to lenses with even more accurate phase-advances, provided enough target functions are tried.

Our specific algorithm for optimization, which takes unit cell transmittance into account in addition to phase advance, is described in detail in the next section; our algorithm for generating target functions is described in Appendix \ref{appen:target_functions}.

%%% _______________ Methods: Selection of optimal lens parameters ________________ %%%
\subsection{Selection of optimal lens parameters}
\label{subsect:lens_parameters}

To choose parameters for each zone of the lens based on the simulation data, the following algorithm was employed:

\begin{enumerate}

\item[1.] Choose a target phase-advance function $\phi_{t_1}(1,f_i)$ (see Appendix \ref{appen:target_functions}) for zone 1. Also choose a target value for transmittance $T_{t}$ for all zones (for best results, transmittance should be uniform throughout all zones; see discussion in Sec.~\ref{subsect:non-ideal-properties}.) After implementing the algorithm using several different values of $T_t$, we found empirically that using $T_t = 0.7$ led to good phase accuracy while still maintaining acceptable levels of transmittance.

\item[2.] Compute a goal function $G$ for every unit cell from the database using the formula

\begin{equation}
\label{eqn:G}
%\begin{split}
G(g,w,\phi_{t_1})  ~ = ~ \sum_i \frac{\left[\delta\phi(g,w,f_i) - \phi_{t_1}(1,f_i) \right]^2}{90}  ~ + ~
\frac{\left[ T(g,w,f_i) - T_{t} \right]^2}{T_t}.
%\end{split}
\end{equation}

Here $\delta\phi(g,w,f_i)$ is the transmitted phase as defined in Sec.~\ref{subsect:database} for the simulated unit cell at frequency $f_i$ and $T(g,w,f_i)$ is its transmittance at $f_i$.

\item[3.] Since a given $\delta\phi$ is equivalent to $\delta\phi + 360^\circ m$ ($m\in\mathbb{Z}$) from the point of view of interference, all unit cells with this equivalence are worthy of consideration for a given zone. Thus compute $G(g,w,\phi_{t_1} + 360^\circ m)$ for $m = 0, -1, -2, ...$ until the sum of $\phi_{t_1}$ and $360^\circ m$ falls more than $100^{\circ}$ below the lowest phase advance measured of all the unit cells. (In practice, all values for $\delta\phi$ computed by the CST solver in the 80--130 GHz band were less than $0^\circ$.)

\item[4.] Select the unit cell that produces the lowest value of $G$.

\item[5.] The target phase-advance function $\phi_{t_1}(1,f_i)$ will specify the target phase-advance function $\phi_{t_1}(n,f_i)$ for all the remaining zones $n$: 

\begin{equation}
\phi_{t_1}(n,f_i) ~ = ~ \phi_{t_1}(1,f_i) ~ + ~ \Delta\phi(n,f_i),
\end{equation}

where the relative phase advance $\Delta\phi(n,f_i)$ is defined in Eq.~\ref{eqn:Delta_phi}. Select the unit cells for which $G$ is lowest in each zone $n$. 

\item[6.] The unit cells selected for each zone in step 5 will form a lens $L_1$. Let $\delta\phi_{L_1}(n,f_i)$ equal the phase-advance of the Zone $n$ unit cell at frequency $f_i$; let $T_{L_1}(n,f_i)$ equal the transmittance of the Zone $n$ unit cell at frequency $f_i$.

\item[7.] Repeat steps 1-6 for a large number $k$ of different target phase-advance functions $\delta\phi_{t_k}(n,f_i)$, which will lead to a set of prototype lenses $L_k$ with transmitted phases $\delta\phi_{L_k}(n,f_i)$ and transmittances $T_{L_k}(n,f_i)$.

\item[8.] Determine the relative phase advances $\Delta\phi_{L_k}$ between the unit cells of each lens $L_k$:

\begin{equation}
\label{eqn:delta_phi_L_k}
\Delta\phi_{L_k}(n,f_i) ~ = ~ \delta\phi_{L_k}(n,f_i) ~ - ~ \delta\phi_{L_k}(1,f_i)
\end{equation}

\item[9.] For each prototype lens $L_k$ compute a ``macro" goal function $M(L_k)$, summed over all frequencies $f_i$ and all zones $n$:

\begin{equation}
\label{eqn:M}
%\begin{split}
M(L_k) ~ = ~ \sum_i \sum_n W_n \frac{ \left[ \Delta\phi_{L_k}(n,f_i) ~ - ~ \Delta\phi(n,f_i) \right]^2 }{90} 
 + ~~W_n \frac{\left[ T_{L_k}(n,f_i) - T_t \right]^2 }{T_t}.
%\end{split}
\end{equation}

Here $\Delta\phi(n,f_i)$ is given in Eq.~\ref{eqn:Delta_phi} and $W_n$ is a weight function (discussed below) given by

\begin{equation}
\label{eqn:weight}
W_n = r_{n} \exp \left(-\frac{r_{n}^2}{s^2} \right),
\end{equation}

where $r_n$ is the annular radius of Zone $n$ and $s$ is the beam radius at the lens. Select the lens $L_*$ for which $M$ is minimal.

\end{enumerate}

The weight function $W_n$ in Eq.~\ref{eqn:weight} is meant to scale the goal function to reflect the relative contribution of each zone $n$ to the coherent sum that determines the electric field amplitude at a given observation point $p$. This contribution is proportional both to the number of unit cells in the $n$th zone ($\propto r_n$) and to the amplitude of the field emitted by the zone's dipoles before taking transmittance into account ($\propto \exp [-{r_n^2}/{s^2} ]$).

The above algorithm owes its complexity to the fact that, strictly speaking, there is no absolute distribution of phase-advances which the zones must match; rather, all that is important is the difference in phase advance between zones (hence the use of \textit{relative} phase advances in $M$). We exploit this flexibility by trying a large number of different target functions $\phi_t(n,f_i)$---each of which leads to the creation of a possible lens $L_k$---and then choosing the lens $L_*$ for which $M$ is minimal. 

%%% ________________ Methods: Further Numerical optimization ____________________ %%%
\subsection{Further numerical optimization}
\label{subsect:numerical_optimization}

The process above identified the ensemble of unit cells from the database (denoted by $L_*$) that best conformed to the desired phase-advances for the lens. These unit cells were further optimized with full time-domain simulations via the CST software. For the unit cell corresponding to the $n^\text{th}$ zone on the lens, the optimization routine would take as inputs: (1) the dimensions $g$ and $w$ of that unit cell and (2) the target function $\phi_{t_*}(n,f_i)$ used to determine $L_*$. The routine then embarked on a series of simulations, each time adjusting the capacitor gaps $g$ and inductor widths $w$ by amounts smaller than the increments used to form the initial database. The routine was also allowed to vary the capacitor gaps in different pairs of layers within the unit cell (\textit{i.e.}, the front and back, 2nd from the front and 2nd from the back, etc.) independently from one another. The goal function $G(\{g\},w,\phi_{t_*})$ (Eqn.~\ref{eqn:G}) was calculated for each simulation, and the routine continued to adjust the dimensions until $M$ (Eqn.~\ref{eqn:M}) converged to a minimum value. (Note that the use of brackets in the argument of $G$ here refers to the fact that the unit cells now have five independently varying $g$ values rather than a uniform $g$ for all capacitive layers).

The purpose of this additional measure was to provide some fine adjustment to the dimensions of the unit cells of $L_*$ to bring their phase-advances even closer to those of the target function $\phi_{t_*}$. The lens consisting of these fully optimized unit cells will be denoted by $L_{**}$.

%%% ___________ Methods: Predicting the performance of the optimized lens ___________ %%%
\subsection{Predicting the performance of the optimized lens}
\label{subsect:predicting_performance}

We then performed computations similar to those in Section \ref{subsect:phase-advances} to compare the focal lengths of the prototype lens $L_{**}$ to our desired focal lengths. Using the dipole array in Fig.~\ref{fig:computation-schematic}, dipoles in the $n$th zone were given an initial phase equal to $\delta\phi_{L_{**}}(n,f_i)$ of the Zone $n$ unit cell at frequency $f_i$. In addition, the dipole amplitudes were multiplied by a factor equal to the square root of the transmittance of the corresponding unit cells.

Focal lengths at each benchmark frequency were computed, as before, by identifying the point at which the coherent sum of the contributions from each dipole to the electric field had the greatest intensity. The beam radius $s(z,f_i)$, or the distance from the propagation axis at which the field amplitude falls to $1/e$ times its value on the axis, was also computed for a range of $z$ values.

%%%  _______________________________ Results ______________________________ %%%
\section{Numerical results}
\label{sect:results}

\paragraph{Notation.} In this section, computations performed for an \textit{ideal lens} are intended to demonstrate the properties of a metamaterial lens with perfect transmittance $T$ and whose unit cells impart precisely the phase-advances prescribed by Eq.~\ref{eqn:phase_shift}; thus calculations are based on the radiation field of an array of electric dipoles (Fig.~\ref{fig:computation-schematic}) with a zoned Gaussian amplitude profile a phase profile determined by Eq.~\ref{eqn:phase_shift}. 

A \textit{simulated lens}, on the other hand, refers to a lens whose unit cells are optimized in the sense of Sec.~\ref{sect:methods}. The amplitudes associated with the different dipoles are Gaussian, but multiplied to the square roots of the simulated transmittances of the respective zones. Phase offsets are determined by the phase-advances of the simulated unit cells.

%%% __________________ Results: Effects of non-ideal lens properties ________________ %%%
\subsection{Finite aperture effects}
\label{subsect:non-ideal-properties}

\begin{figure}
\includegraphics[width=7cm]{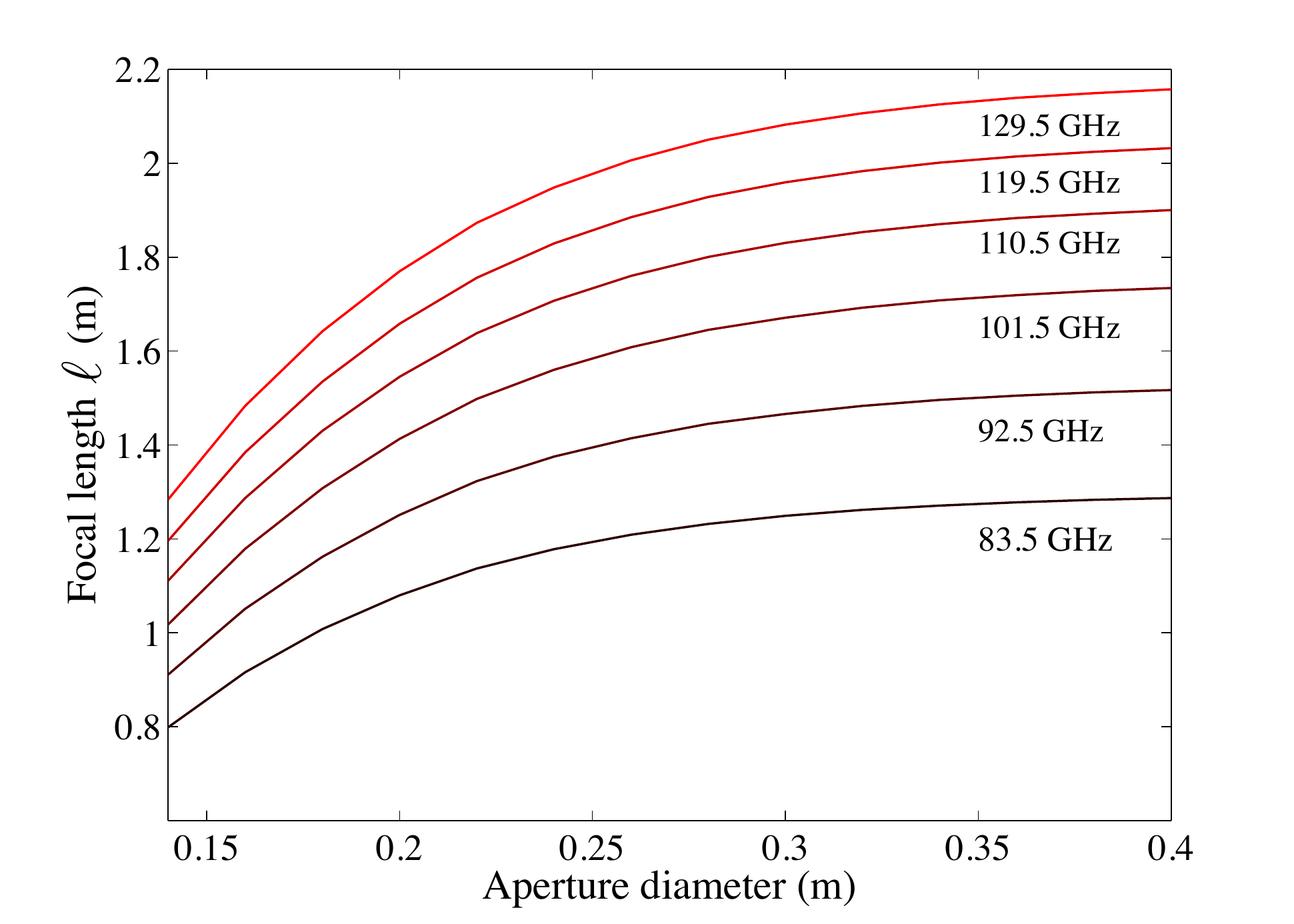}
\caption{Focal lengths of an ideal lens at the benchmark frequencies. The beam radius at the lens is 9.8 cm. The horizontal asymptotes indicate what the focal lengths would be if the aperture were infinite. Note that the maximum diameter for the DIII-D setup is 0.3 m; larger apertures are plotted here for clarity.}
\label{fig:finite-aperture}
\end{figure}

Ideal lens computations showed that the distance of the beam waist from the lens was affected by the size of the aperture relative to the beam radius at the lens plane. Fig.~\ref{fig:finite-aperture} shows that increasing the lens aperture increases the focal length $\ell$ at all frequencies, as well as the spread of $\ell$ with $f$ (essentially, the amount of chromatic aberration).

This effect may be advantageous, especially if combined with a radial repositioning of the lens. If the toroidal magnetic field in DIII-D is strengthened, for example, the EC-emitting locations of the benchmark frequencies will move to smaller radii $R_{\text{maj}}$ and become closer together to one another. The lens may be adapted to the overall movement by being moved itself; furthermore, as these results indicate, the lens may adapt to the change in spacing between the locations by narrowing its aperture (e.g., with a diaphragm). The main drawback of altering the lens aperture would be the reduction in lens transmittance.

%%% ___________________ Results: Lens parameter optimization ___________________ %%%
\subsection{Lens parameter optimization}
\label{subsect:parameter_optimization}

\begin{figure}
\includegraphics[width=12cm]{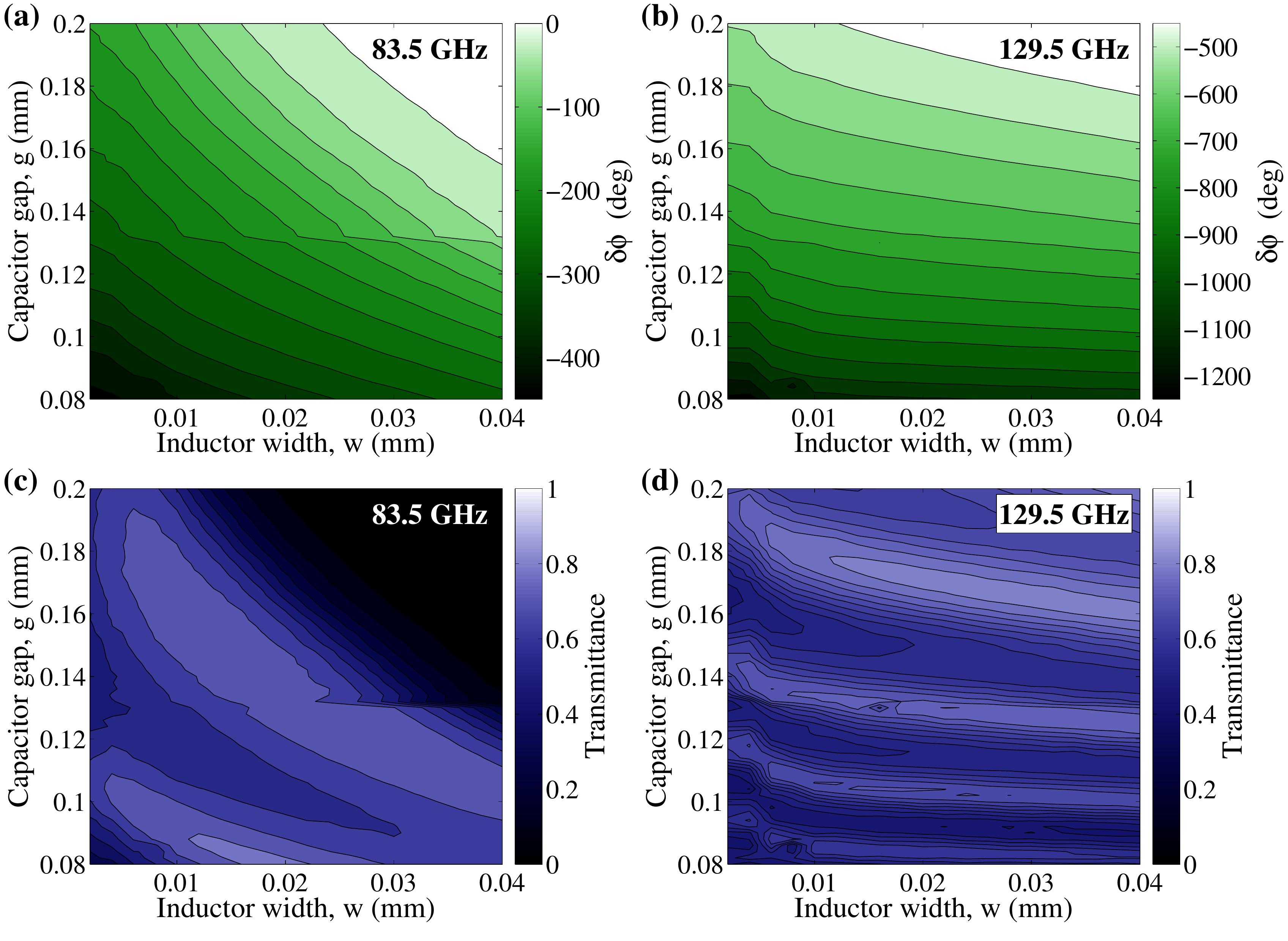}
\caption{Contour plots of the simulated $\delta\phi$ (a,b) and transmittance (c,d) of electromagnetic radiation at the noted frequency through a unit cell with the parameters listed on the axes.}
\label{fig:phase_trans_contours}
\end{figure}

\begin{figure}
\includegraphics[width=12cm]{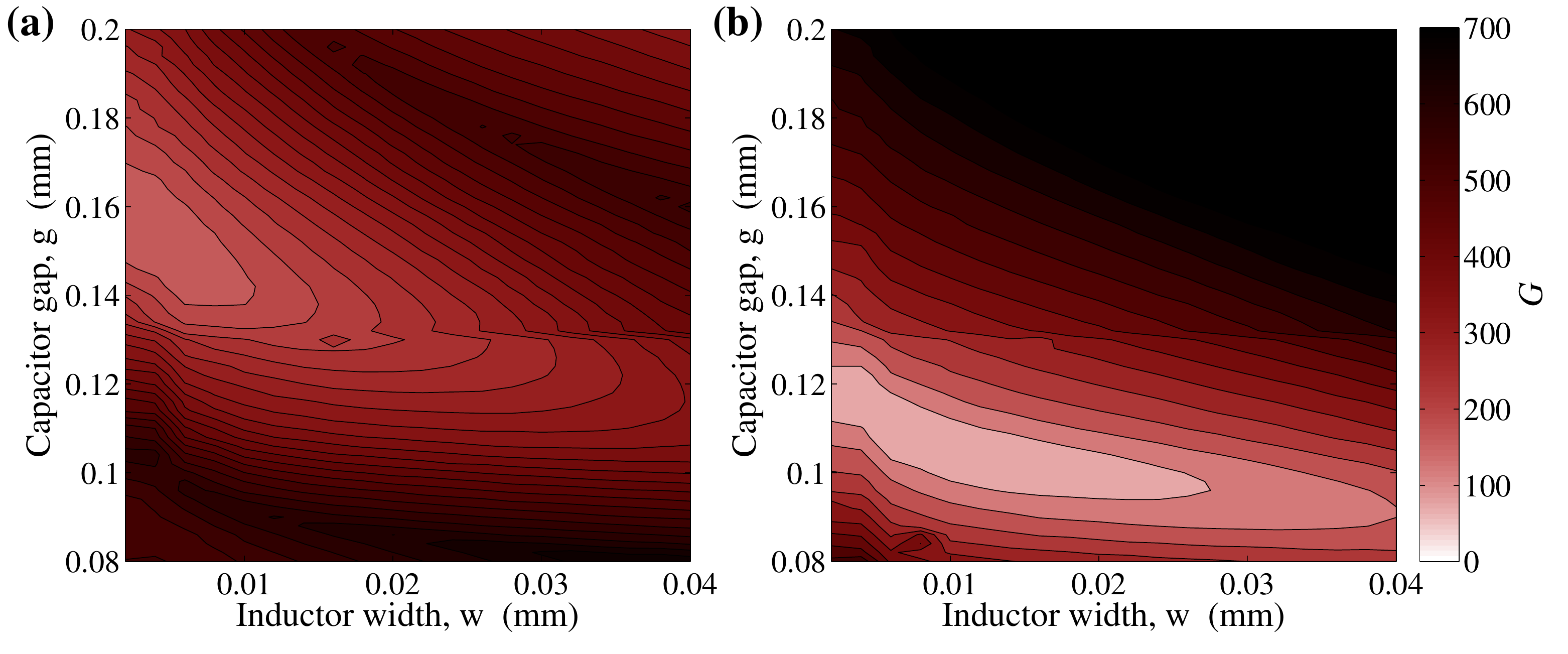}
\caption{Contour plots of the goal function $G(g,w,\phi_{t*})$ for Zone 1 unit cells for (a) the case in which the target functions $\phi_t$ (of which $\phi_{t*}$ leads to the minimal $M$) are restricted to depend linearly on $f$ as in Ref.~\cite{capecchi2012}, and (b) the case in which the target functions $\phi_t$ were chosen as per the method in Appendix \ref{appen:target_functions} (\textit{i.e.}, they are defined quasi-independently at each of the six benchmark frequencies listed in Sec.~\ref{subsect:database}.) Values of $G$ greater than 700 are not differentiated on the color axis.}
\label{fig:comparison}
\end{figure}

The transmittances and phase-advances associated with the unit cell dimensions covered in our database are plotted in Fig.~\ref{fig:phase_trans_contours} for two of the benchmark frequencies. 

Fig.~\ref{fig:comparison} shows the values of $G$ for the database unit cells according to the optimized target function for zone 1 $\phi_{t*}(1,f)$. In particular, in Fig.~\ref{fig:comparison}a the target functions $\phi_{t*}$ were restricted to be linear functions of $f$, and were thus specified by two parameters. This is the same procedure followed in a previous work of ours.\cite{capecchi2012} In Fig.~\ref{fig:comparison}b, by contrast, the target functions were determined as per Appendix \ref{appen:target_functions}. The noticeably smaller values of $G$ reached in Fig.~\ref{fig:comparison}b as opposed to Fig.~\ref{fig:comparison}a confirm that the advantages of the new procedure in more closely adhering to the desired lens phase profile.

\begin{figure}
\includegraphics[width=7cm]{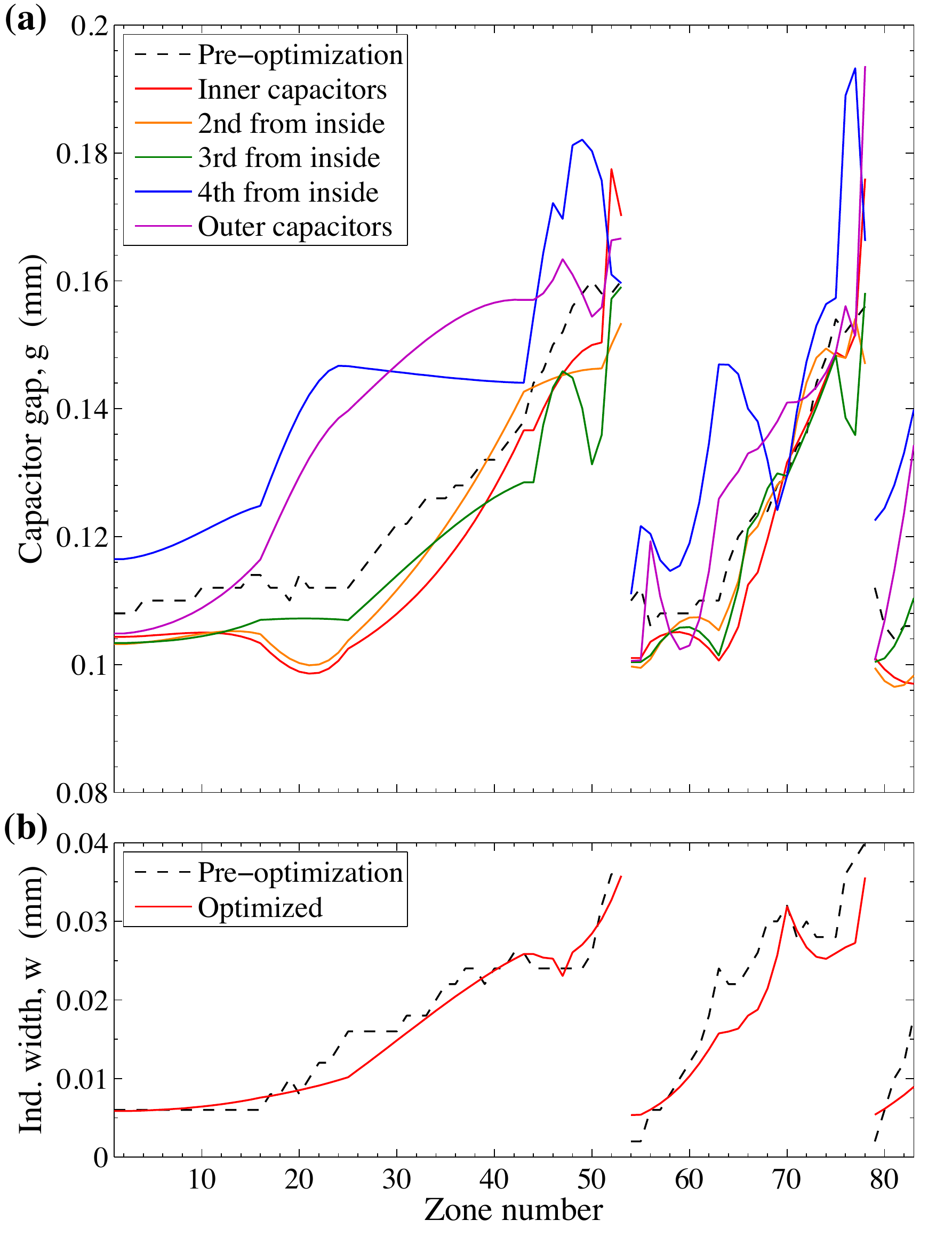}
\caption{Dimensions of the unit cells for each zone as determined by the optimization. (a) Capacitor gap widths $g$ for pairs of capacitive layers indicated by the line color. (b) Inductive widths $w$. Dashed lines in both plots represent the dimensions chosen by the algorithm in Sec.~\ref{subsect:lens_parameters}, which were the starting point for the optimizations described in Sec.~\ref{subsect:numerical_optimization}.}
\label{fig:parameters}
\end{figure}

The final unit cell dimensions for each zone, determined by the fine-grained optimization routine of Sec.~\ref{subsect:numerical_optimization}, are plotted in Fig.~\ref{fig:parameters}. The transmitted phase $\delta\phi$ and transmittance $T$ associated with unit cells of these dimensions were used for the predictions of lens performance in the following subsection.

%%% ___________________ Results: Predicted lens performance ____________________ %%%
\subsection{Predicted lens performance}
\label{subsect:predicted_performance}

\begin{figure}
\includegraphics[width=8cm]{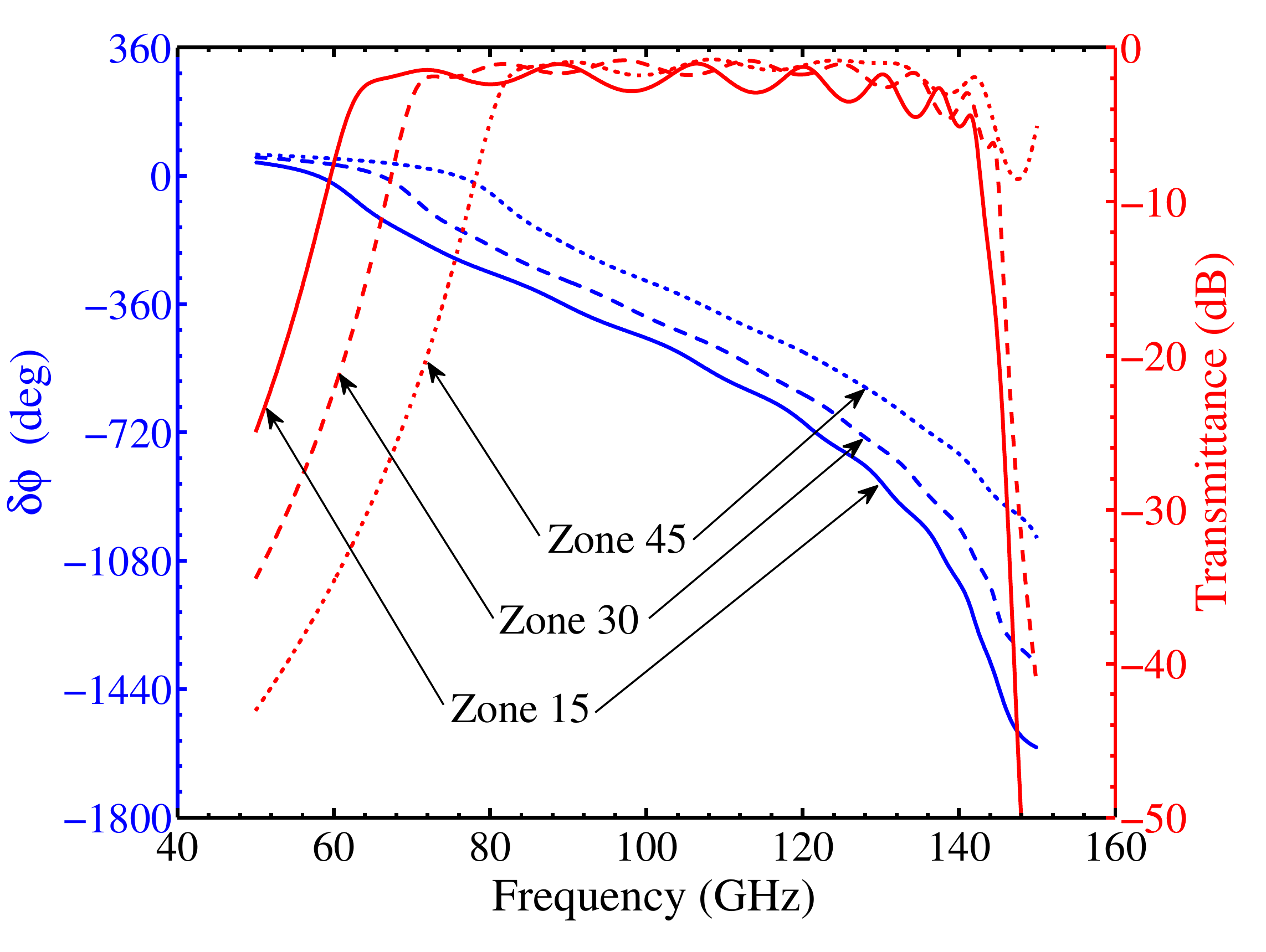}
\caption{Plots of $\delta\phi$ (blue) and transmittance (red) through the unit cells corresponding to zone 15 (solid lines), zone 30 (dashed lines), and zone 45 (dotted lines) of the simulated lens. Note the rightward movement of the pass-band that accompanies the (desired) rightward movement of $\delta\phi$.}
\label{fig:transfer}
\end{figure}

The $\delta\phi$ and transmittance of three exemplary unit cells of the simulated lens are plotted in Fig.~\ref{fig:transfer}. As is required for lens-like behavior, the relative phase advance $\Delta\phi$ (Eqn.~\ref{eqn:Delta_phi}) is greater for the zones that are farther from the symmetry axis. Note, also, the rightward movement of the unit cell pass-band that accompanies the shifts in $\delta\phi(f)$. This movement can enforce limits on the flexibility of the MEFSS-based lens in attaining arbitrary $\ell(f)$ distributions. Specifically, the attainment of some phase profiles may be impossible if certain desired frequencies (83--130 GHz in our case) are excluded from the pass-band in some zones. Indeed, as we discuss below, some of the zones in our simulated lens do exhibit this frequency exclusion.

\begin{figure}
\includegraphics[width=12cm]{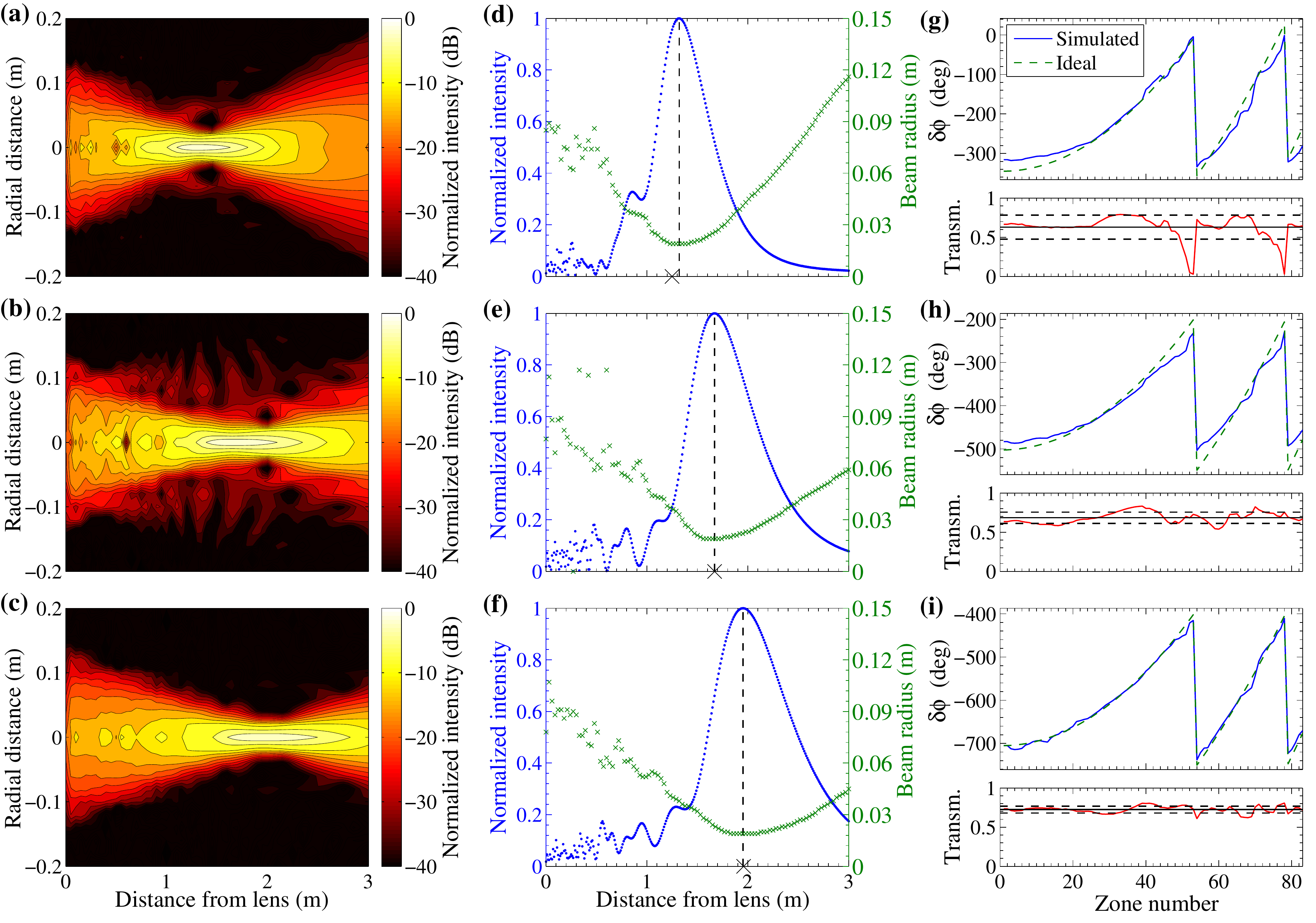}
\caption{(a-c) Contours of field intensity, (d-f) intensity along the central axis normal to the lens and beam radius, and (g-i) $\delta\phi$ (defined in Sec.~\ref{subsect:database}) and transmittance as a function of zone number for simulated lenses at the three benchmark frequencies indicated in (a-c) (and applicable to the respective rows of plots). The dashed vertical lines in (d-f) indicate the points of maximum field intensity (\textit{i.e.}, beam waist locations or focal lengths) for the simulated lenses; $\times$-marks on the $x$-axis indicate the desired beam waist locations. Note that each annular zone is three unit cells wide along the $x$- and $y$-axes of the lens, so one unit on the $x$-axis of (m-r) corresponds approximately with a radial distance of 0.0018 m. Note the forward movement of the beam waist with frequency. }
\label{fig:field-z-rad-trans}
\end{figure}

Plots of the fields, beam radii, phase-advances, and transmittances of simulated lenses at selected frequencies are shown in Fig.~\ref{fig:field-z-rad-trans}. The intensity contours (Figs.~\ref{fig:field-z-rad-trans}a-c) and beam radii (Figs.~\ref{fig:field-z-rad-trans}d-f) bear close resemblance to Gaussian modes as expected, with perturbations---especially in the near field---due to interference effects resulting from deviations of phase advance and transmittance from their desired values (Figs.~\ref{fig:field-z-rad-trans}g-i). The overall trend of reverse chromatic aberration, in which the focal length moves away from the lens with increasing frequency, can be seen in \ref{fig:field-z-rad-trans}a-f.

It is clear from Fig.~\ref{fig:field-z-rad-trans}g-i that our algorithm, and the subsequent CST optimization routine, were unable to construct a lens without significant deviations from the desired profiles of phase-advance and transmittance: note, in particular, the significant deviations from an ideal relative phase profile at both 83.5 GHz and 101.5 GHz (Fig.~\ref{fig:field-z-rad-trans}g,h) and the steep minima in transmittance at 83.5 GHz (Fig.~\ref{fig:field-z-rad-trans}g). The minima in transmittance result from the movement of the pass-band mentioned above, which causes the lower frequencies of interest to be cut off. Further improvements to our algorithm, as well as an expansion of our unit cell database, may lead to reductions in these deviations. However, the deviations may also reflect inherent physical limitations of the MEFSS concept---\textit{i.e.}, it may be impossible for a 10th-order unit cells to achieve the desired ranges of frequency and phase responses necessary for the construction of a lens that matches the ideal characteristics.

The foci of the simulated lens are indicated by the vertical dashed lines in Fig.~\ref{fig:field-z-rad-trans}d-f. These were computed by determining the distance front the lens (along the symmetry axis) at which the field intensity was maximal. The desired focal lengths, as listed in Table \ref{table:benchmark_frequencies}, are shown as $\times$-marks on the $x$-axes of Fig.~\ref{fig:field-z-rad-trans}d-f. Note the relative closeness of the simulated focal lengths to the desired focal lengths despite the deviations in phase and transmittance mentioned above: the largest deviation in focal length among the benchmark frequencies was 7.1 cm at 83.5 GHz (Fig.~\ref{fig:field-z-rad-trans}d)---an error of 5.7\%. 

\begin{figure*}
\includegraphics[width=12cm]{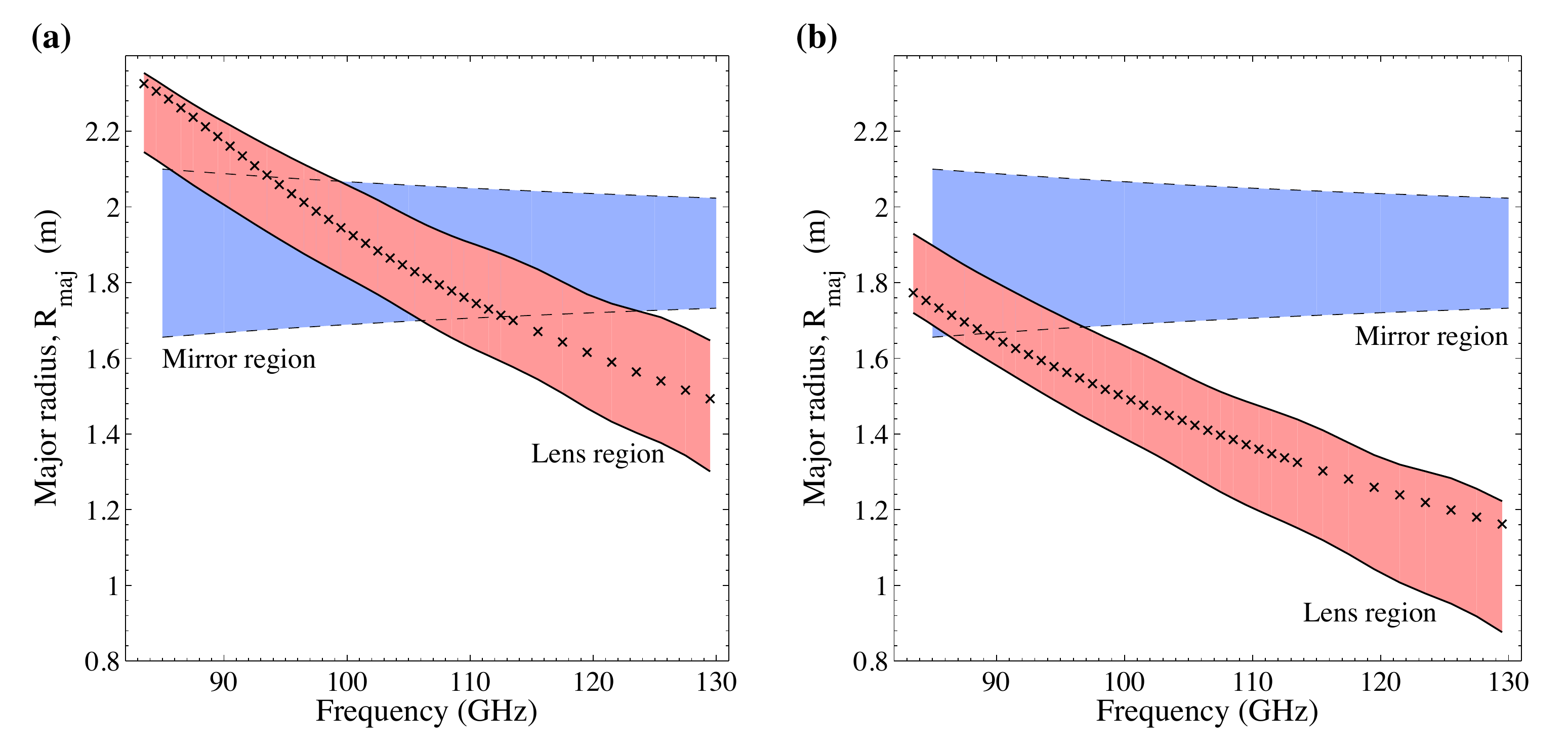}
\caption{EC-emitting locations of selected frequencies (denoted by $\times$-marks) in the DIII-D tokamak, relative to the beam waist regions of the metamaterial lens (the shaded region between the solid lines) and the current ellipsoidal mirror (the shaded region between the dashed lines). In (a), the toroidal magnetic field is -2.00 T and the lens is assumed to be positioned at a major radius $R_\text{maj}$ of 3.575 m; in (b), the field is -1.57 T and the lens is at $R_\text{maj} = 3.150$ m. The aforementioned \textit{beam waist regions} are bounded by the locations at which the beam radius is 5\% greater than the minimum radius of the Gaussian mode associated with the corresponding focusing optic at the frequency indicated on the $x$-axis.}
\label{fig:ece_locations}
\end{figure*}

Fig.~\ref{fig:ece_locations} depicts the relative accuracy of the simulated lens versus the current ellipsoidal mirror as a focusing optic for EC emission in the DIII-D tokamak. As the figure indicates, emission at all frequencies within the range of interest fall within the beam waist region of the lens, both at the maximum (Fig.~\ref{fig:ece_locations}a) and minimum (Fig.~\ref{fig:ece_locations}b) intensities of the confining toroidal magnetic field. We define the \textit{beam waist regions} as the regions in which the beam radius of the Gaussian mode associated with the relevant focusing optic is with 5\% of its minimum beam-waist value at a given frequency. Note that, whereas the location of the ellipsoidal mirror is fixed, we have exploited the planned translational freedom of the metamaterial lens along the major radial direction. (The lens aperture diameter, however, was assumed to be the 30 cm in both cases.)

%%% ____________________________ Conclusions _____________________________ %%%
\section{Conclusions}
\label{sect:conclusions}

We have presented the design and numerical optimization of a metamaterial lens exhibiting reverse chromatic aberration---that is, a focal length increasing with frequency---in the 83-130 GHz range. This behavior is not encountered in any convergent lens made of natural materials. It could, in principle, be obtained by a generalization of the achromatic doublet, but would suffer from the practical limitations of arraying several lenses of finite thickness, whereas the metamaterial lens proposed has a thickness comparable with or smaller than the wavelength. 

The lens has been optimized according to a new numerical method presented here in Sec.~\ref{subsect:lens_parameters}. Significant improvements over a previous method \cite{capecchi2012} are exemplified by Fig.~\ref{fig:comparison}, indicating a better agreement between the actual and the desired frequency dependence of the phase-advance and transmittance. Although this agreement is not perfect (Fig.~\ref{fig:field-z-rad-trans}), our calculations indicate that the location of beam waist at each frequency  of interest is sufficiently accurate (Fig.~\ref{fig:ece_locations}) for the lens to be a marked improvement over the ellipsoidal mirror that the lens is intended to replace.

The lens was optimized for possible deployment with the Electron Cyclotron Emission radiometer in the DIII-D tokamak. We predict (Fig.~\ref{fig:ece_locations}) that it will correctly and simultaneously focus different frequencies to their respective emitting locations, which are separated by up to 0.85 m. For reference, the tokamak has a radius R = 1.66 m and the lens would be located at a minimum $R_\text{maj}$ of 3.15 m. We also show that a simple movement of the lens can compensate for displacements of the emitting locations caused by changes to the magnetic field. Because ECE at DIII-D undergoes one of the largest variations of optimal focal length with frequency, the results presented are encouraging in that similar metamaterial lenses can be designed for other millimeter wave diagnostics and/or devices.

Although the primary objective of the work described in the paper was to design an MEFSS-based lens for ECE detection on the DIII-D tokamak, our optimization techniques may be applied more broadly to the design of metamaterial gradient-index lenses. In particular, the algorithm is not limited to the design of lenses with reverse-chromatic aberration. It can generate a lens optimized for any desired frequency dependence of focal length $\ell(f)$ as specified by the desired profile of relative phase-advance, derived in Sec.~\ref{subsect:phase-advances}. The only required input would be data on transmittance and phase-advance for a variety of simulated unit cells. This procedure may, therefore, be applied to the design of any metamaterial lens consisting of discrete unit cells for which radiation and transfer properties have been computed in advance through simulation.

\appendix

\section{Choosing target functions}
\label{appen:target_functions}

The target functions $\phi_t(n,f)$ were chosen according to the following steps:

\begin{enumerate}

\item[1.] For every unique pair $\{a,b\}$ database unit cells, compute the difference in $\delta\phi$ at each of the benchmark frequencies; in other words, find $\delta\phi_b(f_i) - \delta\phi_a(f_i)$.

\item[2.] Compute the following sum over the six benchmark frequencies $f_i$ listed in section \ref{subsect:database}:

\begin{equation}
\chi_{\{a,b\}}^2 = \sum_i \left[ \left(\delta\phi_b(f_i) - \delta\phi_a(f_i) \right) - \Delta\phi(83,f_i) \right]^2
\end{equation}

where $\Delta\phi$, defined in Eqn.~\ref{eqn:Delta_phi}, signifies the difference in phase advance between the outermost zone (83) and the innermost zone (1) at frequency $f_i$.

\item[3.] Identify the pair $\{a^*,b^*\}$ for which $\chi^2$ is the lowest. Choose as the initial target phase distribution the following:

\begin{equation}
\phi_{t_1}(1,f_i) = \delta\phi_{a^*}(f_i) + \frac{[\delta\phi_{b^*}(f_i)-\delta\phi_{a^*}(f_i)] - \Delta\phi(83,f_i)}{2}
\end{equation}

\item[4.] Use as the remaining target phase distributions $\phi_{t_k}$ the set:

\begin{equation}
\phi_{t_k}(1,f) = \begin{cases}
\phi_{t_1}(1,f_1) + \theta_1 &\mbox{if } f = f_1 \\
\phi_{t_1}(1,f_2) + \theta_2 &\mbox{if } f = f_2 \\
\phi_{t_1}(1,f_3) + \theta_3 &\mbox{if } f = f_3 \\
\phi_{t_1}(1,f_4) + \theta_4 &\mbox{if } f = f_4 \\
\phi_{t_1}(1,f_5) + \theta_5 &\mbox{if } f = f_5 \\
\phi_{t_1}(1,f_6) + \theta_6 &\mbox{if } f = f_6
\end{cases}
\end{equation}

where $\theta_1,\theta_2,\theta_3,\theta_4,\theta_5,\theta_6$ can take the values $0, \pm 4^{\circ},$ and $\pm 8^{\circ}$. This will lead to $6^5$ unique $\phi_{t_k}$ functions.

\end{enumerate}


\end{document}